\RequirePackage{fix-cm}
\documentclass[twocolumn]{svjour3}
\usepackage[sort,compress,numbers]{natbib}
\smartqed  
\usepackage{hhline}
\usepackage[T1]{fontenc}
\usepackage{graphicx}
\usepackage{newtxtext,newtxmath}
\usepackage{flushend}
\usepackage[colorlinks,citecolor=blue,urlcolor=blue,linkcolor=blue]{hyperref}
\usepackage[utf8]{inputenc}
\usepackage{amsmath}
\usepackage{mathtools}
\usepackage{xspace}
\newcommand{\pt}{\ensuremath{p_{\mathrm{T}}}\xspace}
\newcommand{\GeV}{\ensuremath{\,\text{Ge\hspace{-.08em}V}}\xspace}

\newcommand{\nnodes}{\ensuremath{n_\mathrm{nodes}\xspace}}
\newcommand{\nedges}{\ensuremath{n_\mathrm{edges}\xspace}}

\DeclareMathAlphabet{\mathcal}{OMS}{cmsy}{m}{n}

\journalname{Computing and Software for Big Science}

\begin{document}
\sloppy
\title{Charged particle tracking via edge-classifying interaction networks

\thanks{S.~T. and V.~R. are supported by IRIS-HEP through the U.S. National Science Foundation (NSF) under Cooperative Agreement OAC-1836650.
J.~D. is supported by the U.S. Department of Energy (DOE), Office of Science, Office of High Energy Physics Early Career Research program under Award No. DE-SC0021187.
G.~D. is supported by DOE Award No. DE‐SC0007968.}
}

\author{Gage DeZoort$^{1}$ \and
Savannah Thais$^{1}$ \and
Javier Duarte$^{2}$\and
Vesal Razavimaleki$^{2}$ \and
Markus Atkinson$^{3}$ \and
Isobel Ojalvo$^{1}$ \and
Mark Neubauer$^{3}$ \and
Peter Elmer$^{1}$
}


\authorrunning{G. DeZoort et al.} 

\institute{G. DeZoort \at
              \email{jdezoort@princeton.edu}
             \at
           S. Thais \at
              \email{sthais@princeton.edu}
              \at
              $^{1}$Princeton University, Princeton, NJ, USA
              \at
              $^{2}$University of California San Diego, La Jolla, CA, USA
              \at
              $^{3}$University of Illinois at Urbana-Champaign, Champaign, IL, USA
}

\date{Received: July 12, 2021 / Accepted: October 13, 2021 / Published: November 15, 2021}

\maketitle

\begin{abstract}
  Recent work has demonstrated that geometric deep learning methods such as graph neural networks (GNNs) are well suited to address a variety of reconstruction problems in high energy particle physics. 
  In particular, particle tracking data is naturally represented as a graph by identifying silicon tracker hits as nodes and particle trajectories as edges; given a set of hypothesized edges, edge-classifying GNNs identify those corresponding to real particle trajectories. 
  In this work, we adapt the physics-motivated interaction network (IN) GNN toward the problem of particle tracking in pileup conditions similar to those expected at the high-luminosity Large Hadron Collider.
  Assuming idealized hit filtering at various particle momenta thresholds, we demonstrate the IN's excellent edge-classification accuracy and tracking efficiency through a suite of measurements at each stage of GNN-based tracking: graph construction, edge classification, and track building. 
  The proposed IN architecture is substantially smaller than previously studied GNN tracking architectures; this is particularly promising as a reduction in size is critical for enabling GNN-based tracking in constrained computing environments. 
  Furthermore, the IN may be represented as either a set of explicit matrix operations or a message passing GNN. 
  Efforts are underway to accelerate each representation via heterogeneous computing resources towards both high-level and low-latency triggering applications. 
\keywords{Graph neural networks \and tracking \and particle physics}
\end{abstract}

\section{Introduction}
\label{intro}

Charged particle tracking is essential to many physics reconstruction tasks including vertex finding~\cite{Piacquadio:2008zzb,cms_tracking}, particle reconstruction~\cite{Aaboud:2017aca,Sirunyan:2017ulk}, and jet flavor tagging~\cite{Larkoski:2017jix,Sirunyan:2017ezt,Aaboud:2018xwy}. 
Current tracking algorithms at the CERN Large Hadron Collider (LHC) experiments~\cite{cms_tracking,atlas_tracking} are typically based on the combinatorial Kalman filter~\cite{combkalman1,combkalman2,combkalman3,kalman} and have been shown to scale worse than linearly with increasing beam intensity and detector occupancy~\cite{hllhc_tracking}. 
The high-luminosity phase of the LHC (HL-LHC) will see an order of magnitude increase in luminosity~\cite{ApollinariG.:2017ojx},
highlighting the need to develop new tracking algorithms demonstrating reduced latency and improved performance in high-pileup environments. 
To this end, ongoing research focuses on both accelerating current tracking algorithms via parallelization or dedicated hardware and developing new tracking algorithms based on machine learning (ML) techniques.

Geometric deep learning (GDL)~\cite{gdl,zhang2020deep,zhou2019graph,wu2019comprehensive} is a growing sub-field of ML focused on learning representations on non-Euclidean domains such as sets, graphs, and manifolds. 
Graph neural networks (GNNs)~\cite{gnnmodel,pointnet,gilmer2017neural,IN,relational,DGCNN} are the subset of GDL algorithms that operate on graphs, data represented as a set of nodes connected by edges, and have been explored for a variety of tasks in high energy physics~\cite{duarte2020graph,Shlomi_2021}. 
Particle tracking data is naturally represented as a graph; detector hits form a 3D point cloud and the edges between them represent hypotheses about particle trajectories. 
Recent progress by the Exa.TrkX project and other collaborations has demonstrated that edge-classifying GNNs are well suited to particle tracking applications~\cite{heptrkx,exatrkx,IN_fpga,Ju:2021ayy}. 
Tracking via edge classification typically involves three stages. 
In the graph construction stage, silicon tracker hits are mapped to nodes and an edge-assignment algorithm forms edges between certain nodes. 
In the edge classification stage, an edge-classifying GNN infers the probability that each edge corresponds to a true track segment meaning that both hits (nodes connecting the edge) are associated to the same truth particle, as discussed further in Sec.~\ref{sec:TrackML}.
Finally, in the track building step, a track-building algorithm leverages the edge weights to form full track candidates.

In this work, we present a suite of measurements at each of these stages, exploring a range of strategies and algorithms to facilitate GNN-based tracking. 
We focus in particular on the interaction network (IN)~\cite{IN}, a GNN architecture frequently used as a building block in more complicated architectures~\cite{heptrkx,IN_fpga,Moreno:2019neq,Moreno:2019bmu}. 
The IN itself demonstrates powerful edge-classification capability and its mathematical formulations are the subject of ongoing acceleration studies~\cite{fpga}. 
In Section~\ref{sec:theory}, we first present an overview of particle tracking and graph-based representations of track hits.
In Section~\ref{sec:IN}, we introduce INs and describe the mathematical foundations of our architecture. 
In Section~\ref{sec:measurements}, we present specific graph construction, IN edge classification, and track building measurements on the open-source TrackML dataset. 
Additionally, we present IN inference time measurements, framing this work in the context of ongoing GNN acceleration studies.  
In Section~\ref{sec:summary}, we summarize the results of our studies and contextualize them in the broader space of ML-based particle tracking. 
We conclude in the same section with outlook and discussion of future studies, in particular highlighting efforts to accelerate INs via heterogeneous computing resources. 

\section{Theory and Background }
\label{sec:theory}

\subsection{Particle Tracking} 
\label{sec-2.1}

\begin{figure*}[!htbp]
\centering
\includegraphics[width=0.8\textwidth,clip]{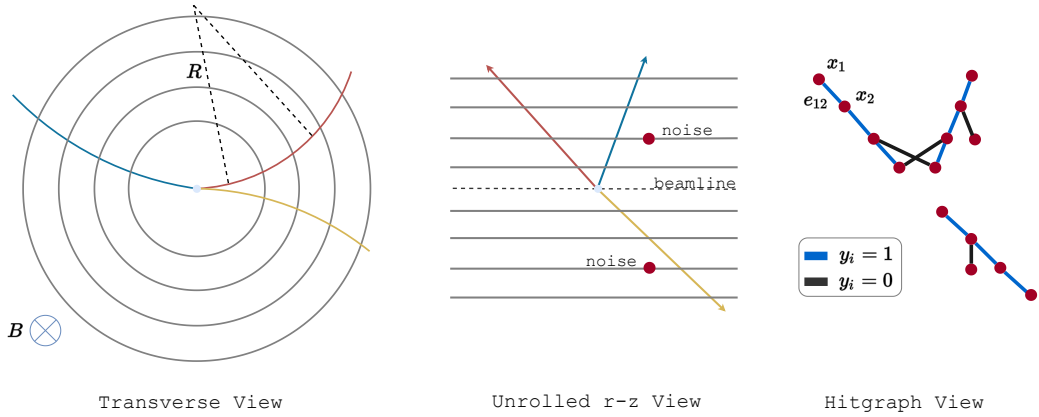}
\caption{(Left) A transverse view of a generic particle tracker, where the $z$-axis points out of the page. 
Here, we see a set of four cylindrical detector layers with three particles traversing them. 
The magnetic field (of strength $B$) is aligned with the $z$-axis such that tracks move with a radius of curvature $R$ in the transverse plane, yielding measurements of transverse momentum via $\pt=0.3\ [\frac{\GeV}{\text{T}\cdot\text{m}}]\ BR$. 
(Middle) The four cylindrical tracker layers are ``unrolled'' in the $r$--$z$ plane to show the full event contents: three particles plus additional noise hits. 
(Right) The corresponding hitgraph is shown with example node and edge labels. }
\label{fig:tracking} 
\end{figure*}

In collider experiments such as the LHC, charged particle trackers are comprised of cylindrical detector layers immersed in an axially-aligned magnetic field. 
The detector geometry is naturally described by cylindrical coordinates $(r,\phi,z)$, where the $z$ axis is aligned with the beamline. 
Pseudorapidity is a measure of angle with respect to the beamline, defined as $\eta\coloneqq-\log\tan\frac{\theta}{2}$ where $\theta$ is the polar angle.
Charged particles produced in collision events move in helical trajectories through the magnetic field, generating localized hits in the tracker layers via ionization energy deposits. 
Track reconstruction consists of ``connecting the dots,'' wherein hits are systematically grouped to form charged particle trajectories. 
We refer to a pair of hits that belong to the same particle as a track segment, such that the line extending between the hits is a linear approximation of the particle's trajectory. 
Note that in a high-pileup scenario, track hits might correspond to multiple overlapping particle trajectories. 
Reconstructed tracks are defined by their respective hit patterns and kinematic properties, which are extracted from each track's helix parameters. 
Specifically, initial position and direction follow directly from helical fits and the transverse momentum $\pt$ is extracted from the track's curvature (see Figure~\ref{fig:tracking})~\cite{tracking}. 

\begin{figure}[!htbp]
\centering
\includegraphics[width=0.99\columnwidth,clip]{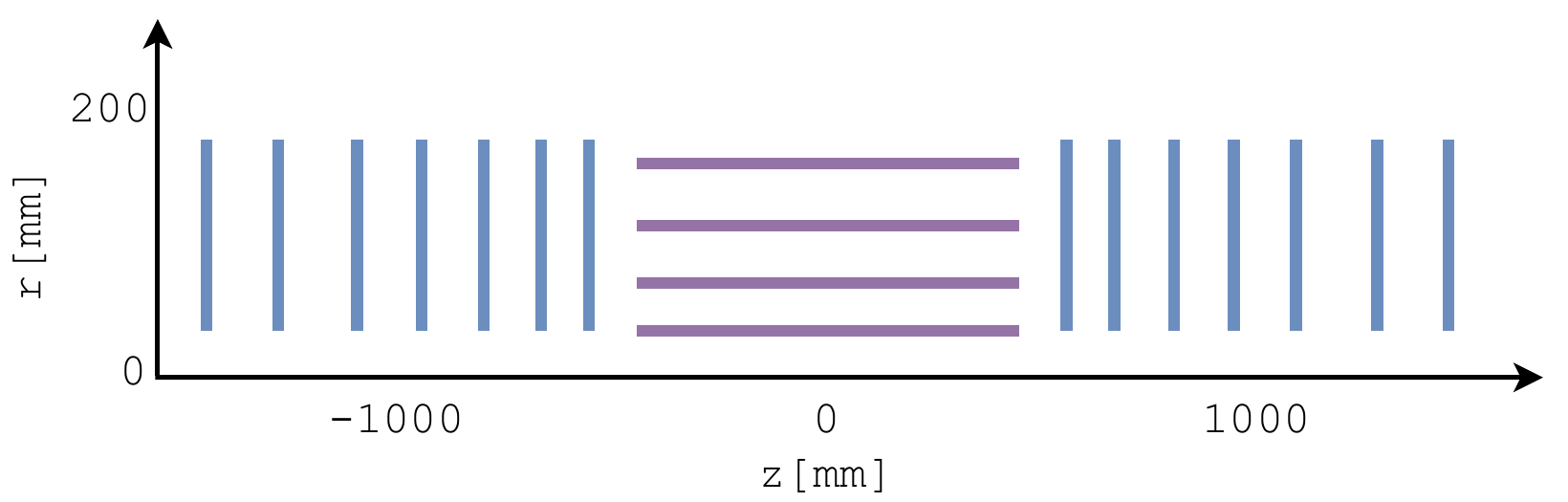}
\caption{Here we depict a particle tracker geometry similar to tracker designs proposed for the HL-LHC era. 
This ``generic tracker'' geometry is used in the TrackML dataset (see Section~\ref{sec:measurements}). 
The generic tracker is comprised of three sub-detectors named by the shape of their silicon modules: the pixel detector, the short strip detector, and the long strip detector. 
Each sub-detector is divided into volumes (numbered 7-18); each volume contains a set of detector layers. 
Volumes 8, 13, and 17 above are referred to as the \textit{barrel} of the detector because their layers sit at a constant cylindrical radius. 
Volumes 7, 9, 12, 14, 16, and 18 comprise the tracker's \textit{endcaps} because of their disk-like shape; endcap layers are positioned at a single point along the $z$-axis. 
The above figure is adapted from a figure in Ref.~\cite{TrackML} and the TrackML detector diagram accompanying Kaggle's TrackML dataset.}
\label{fig:pixel} 
\end{figure}

In this work, we focus specifically on track building in the pixel detector (see Fig.~\ref{fig:pixel}), the innermost subdetector of the tracker. 
Many tracking algorithms run ``inside out,'' where track seeds from the pixel detector are used to estimate initial track parameters and propagated through the full detector \cite{cms_tracking}.  
Improving the seeding stage of the tracking pipeline is an important step towards enabling efficient tracking at the HL-LHC; this approach is complimentary to other GNN-based tracking efforts that focus on the full tracker barrel (without including endcaps) using graph segmentation \cite{exatrkx}. 

\subsection{Tracker Hits as Graphs}
Tracking data is naturally represented as a graph by identifying hits as nodes and track segments as (in general) directed edges (see Fig.~\ref{fig:tracking}). 
In this scheme, nodes have cylindrical spatial features $x_k=(r_k,\phi_k,z_k)$ and edges are defined by the nodes they connect.
We employ two different edge representations: 1) binary incidence matrices $R_i, R_o\in\{0,1\}^{\nedges\times \nnodes}$ in incoming/outgoing (IO) format and 2) hit index pair lists $I\in\mathbb{N}^{2\times n_{edges}}$ in coordinate (COO) format \cite{fey2019fast}.
Specifically, the incidence matrix elements $(R_i)_{e,h}$ are $1$ if edge $e$ is incoming to hit $h$ and $0$ otherwise; $R_o$ is defined similarly for outgoing edges. 
COO entries $I_{0,e}$ and $I_{1,e}$ are the hit indices from which edge $e$ is outgoing from and incoming to respectively.
Each edge is assigned a set of geometric features $a_{ij} = (\Delta r_{ij}, \Delta \phi_{ij}, \Delta z_{ij}, \Delta R_{ij})$, where $\Delta R_{ij}=\sqrt{\Delta\eta_{ij}^2 + \Delta\phi_{ij}^2}$ is the edge length in $\eta$-$\phi$ space. 
Node and edge features are stacked into matrices $X = [x_k]\in\mathbb{R}^{\nnodes\times 3}$ and $R_a=[a_{ij}]\in\mathbb R^{\nedges\times 4}$. 
Accordingly, we define \textit{hitgraphs} representing tracking data as $\mathcal{G}_\mathrm{IO}\coloneqq (X, R_a, R_i, R_o)$ and $\mathcal{G}_\mathrm{COO}\coloneqq (X,R_a,I)$. 
The corresponding training target is the vector $y\in\mathbb{R}^{\nedges}$, whose components $y_e$ are $1$ when edge $e$ connects two hits associated to the same particle and $0$ otherwise.

\section{Interaction Networks}
\label{sec:IN}

The IN is a physics-motivated GNN capable of reasoning about objects and their relations~\cite{IN}. 
Each IN forward-pass involves a relational reasoning step, in which an interaction is computed, and an object reasoning step, in which interaction effects are aggregated and object dynamics are applied. 
The resulting predictions have been shown to generate next-timestep dynamics consistent with various physical principles. 
We adapt the IN to the problem of edge classification by conceptualizing each hitgraph as a complex network of hit ``objects'' and edge ``relations.''
In this context, the relational and object reasoning steps correspond to edge and node re-embeddings respectively. 
In an edge classification scheme, the IN must determine whether or not each edge represents a track segment. 
Accordingly, we extend the IN forward pass to include an additional relational reasoning step, which produces an edge weight for each edge in the hitgraph. 
We consider two formulations of the IN: (1) the matrix formulation, suitable for edge-classification on $\mathcal{G}_\mathrm{IO}$ defined via \textsc{PyTorch}~\cite{pytorch} and (2) the message passing formulation, suitable for edge-classification on $\mathcal{G}_\mathrm{COO}$ defined via \textsc{PyTorch Geometric} (\textsc{PyG}) ~\cite{fey2019fast}. 
These formulations are equivalent in theory, but specific implementations and training procedures can vary their computational and physics performance. 
In particular, the COO encoding of the edge adjacency can greatly reduce the memory footprint for training. 
For this reason, the measurements performed in this paper are based on the message passing formulation. 
In Section~\ref{matrixformulation}, we review the matrix formulation as presented in the original IN paper \cite{IN}, subsequently expanding the notation to describe the message passing IN formulation in Section ~\ref{messagepassingformulation}. 

\subsection{Matrix Formulation}\label{matrixformulation}
The original IN was formulated using simple matrix operations interpreted as a set of physical interactions and effects \cite{IN}. 
The forward pass begins with an input hitgraph $\mathcal{G}_{IO}=(X, R_a, R_i, R_o)$.
The hits receiving an incoming edge are given by $X_i \coloneqq R_iX \in\mathbb{R}^{ \nedges\times 3}$; likewise, the hits sending an outgoing edge are given by $X_o \coloneqq R_oX \in\mathbb{R}^{\nedges\times 3}$. 
Interaction terms are defined by the concatenation $m(\mathcal{G}_{IO})\coloneqq [X_i, X_o, R_a] \in \mathbb{R}^{ \nedges\times 10}$, known as the marshalling step. 
A relational network $\phi_{R,1}$ predicts an effect for each interaction term, $E\coloneqq \phi_{R,1}\big(m(\mathcal{G}_{IO})\big)\in\mathbb{R}^{\nedges\times 4}$. 
These effects are aggregated via summation for each receiving node, $A\coloneqq a(\mathcal{G}_{IO}, E) = R_i^TE\in \mathbb{R}^{\nnodes\times 4}$, and concatenated with $X$ to form a set of expanded hit features $C \coloneqq [X, A]\in\mathbb{R}^{\nnodes\times 7}$. 
An object network $\phi_O$ re-embeds the hit positions as $\tilde{X} \coloneqq \phi_O(C) \in \mathbb{R}^{\nnodes\times 3}$. 
At this point, the traditional IN inference is complete, having re-embedded both the edges and nodes. 
Accordingly, we denote the re-embedded graph $\mathrm{IN}(\mathcal{G}_{IO})=\tilde{\mathcal{G}}_{IO}=(\tilde{X}, E, R_i, R_o)$.

In order to produce edge weights, an additional relational reasoning step is performed on $\tilde{\mathcal{G}}_{IO}$. 
Re-marshalling yields new interaction terms $m(\tilde{\mathcal{G}}_{IO}) = [\tilde{X}_i, \tilde{X}_o, E]\in\mathbb{R}^{ \nedges\times 10}$ and a second relational network $\phi_{R,2}$ predicts edge weights for each edge: $W(\mathcal{G}_{IO})\coloneqq \phi_{R,2}(m(\tilde{\mathcal{G}}_{IO}))\in(0,1)^{\nedges}$. 
Summarily, we have a full forward pass of the edge classification IN: 
\begin{align}
    W(\mathcal{G}_{IO}) &= \phi_{R,2}\bigg[ m\big(\mathrm{IN}(\mathcal{G}_{IO})\big)\bigg] 
\end{align}

\subsection{Message Passing Formulation}\label{messagepassingformulation}
The message passing NN (MPNN) framework summarizes the behavior of a range of GNN architectures including the IN \cite{gilmer2017neural}. 
In general, MPNNs update node features by aggregating ``messages,'' localized information derived from the node's neighborhood, and propagating them throughout the graph. 
This process is iterative; given a message passing time $T$ indexed by $t\in\mathbb{N}$, a generic message passing node update can be written as follows: 
\begin{align}
    x^{(t)}_{i} = \phi_\mathrm{node}^{(t)}\bigg( x_i^{(t-1)}, \underset{j\in N(i)}{\square} \phi_\mathrm{message}^{(t)}\big(x_i^{(t-1)}, x_j^{(t-1)}, a_{ij}^{(t-1)} \big)  \bigg)
\end{align}
Here, $N(i)$ is neighborhood of node $i$. 
The differentiable function $\phi_\mathrm{message}^{(t)}$ calculates messages for each $j\in N(i)$, which are aggregated across $N(i)$ by a permutation-invariant function $\square$. 
A separate differentiable function $\phi_\mathrm{node}^{(t)}$ leverages the aggregated messages to update the node's features. 
Given this generalized MPNN, the IN follows from the identifications $\phi_\mathrm{message}\rightarrow \phi_{R,1}$, $\square_{j\in N(i)}\rightarrow \sum_{j\in N(i)}$, and $\phi_\mathrm{node}\rightarrow \phi_{O}$ for a single timestep $(T=1)$:
\begin{align}
    a_{ij}^{(1)} &= \phi_{R,1}\big(x_i^{(0)}, x_j^{(0)}, a_{ij}^{(0)} \big)\label{edgeupdate}\\
    x^{(1)}_{i} &= \phi_{O}\bigg(x_i^{(0)}, \underset{j\in N(i)}{\sum}a_{ij}^{(1)} \bigg)\label{nodeupdate}
\end{align}
An additional relational reasoning step gives edge weights
\begin{align}
    w_{ij}^{(1)} \coloneqq \phi_{R,2}\big(x_i^{(1)}, x_j^{(1)}, a_{ij}^{(1)}\big)\label{edgeweight}
\end{align}
In this way, we produce edge weights $W(\mathcal{G}_\mathrm{COO})=[w_{ij}^{(1)}]$ from the re-embedded graph with node features $\tilde{X}=[x_i^{(1)}]$ and edge features $E=[a_{ij}^{(1)}]$. 
This formulation is easily generalized to $T>1$ by applying Eqns.~\ref{edgeupdate} and ~\ref{nodeupdate} in sequence at each time step before finally calculating edge weights via Eqn.~\ref{edgeweight} at time $T$. 
In the following studies, we focus on the simplest case of nearest-neighbor message passing ($T=1$). 

\section{Measurements}
\label{sec:measurements}

\subsection{TrackML Dataset}
\label{sec:TrackML}
The TrackML dataset is a simulated set of proton-proton collision events originally developed for the TrackML Particle Tracking Challenge~\cite{TrackML}. 
TrackML events are generated with 200 pileup interactions on average, simulating the high-pileup conditions expected at the HL-LHC. 
Each event contains 3D hit position and truth information about the particles that generated them. 
In particular, particles are specified by particle IDs ($p_\mathrm{ID}$) and three-momentum vectors ($\mathbf{p}$).
Each simulated hit has a unique identifier assigned that gives the true hit position and which particle created the hit.
For this truth assignment, no merging of reconstructed hits is considered as merging of hits occurs in less
than 0.5\% of the cases and the added complexity was deemed unnecessary for the original challenge.
Other simplifications in this dataset include a simple geometry with modules arranged in cylinders and disks, instead of a more complex geometry with cones, no simulation of electronics, cooling tubes, and cables, and only one type of physics process (top quark-antiquark pairs) instead of a variety of processes.

The TrackML detector is designed as a generalized LHC tracker; it contains discrete layers of sensor arrays immersed in a strong magnetic field. 
We focus specifically on the pixel layers, a highly-granular set of four barrel and fourteen endcap layers in the innermost tracker regions. 
The pixel layers are shown in Figure~\ref{fig:pixel}.
We note that constraining our studies to the pixel layers reduces the size of the hitgraphs such that they can be held in memory and processed by the GNN without segmentation.

\subsection{Graph Construction}
\label{sec:Graph}
In the graph construction stage, each event's tracker hits are converted to a hitgraph through an edge selection algorithm. 
Typically, a set of truth filters are applied to hits before they are assigned to graph nodes. 
For example, \textit{$\pt$ filters} reject hits generated by particles with $\pt<\pt^\mathrm{min}$, \textit{noise filters} reject noise hits, and \textit{same-layer filters} reject all but one hit per layer for each particle. 
These truth filters are used to modulate the number of hits present in each hit graph to make it more feasible to apply GNN methods and can be thought of as an idealized hit filtering step (see Table~\ref{tab:hit_losses}).
One goal of future R\&D is to lower or remove this truth-based filter or replace it with a realistic hit filtering step that could be applied in a high-pileup experimental setting.
After initial hit filtering yields a set of nodes, edge-assignment algorithms extend edges between certain nodes. 
These edges are inputs to the inference stage and must therefore represent as many true track segments as possible. 
Naively, one might return a fully-connected hitgraph. 
However, this strategy yields $\frac{1}{2}\nnodes(\nnodes-1)$ edges, which for $\nnodes\sim\mathcal{O}(1,000)$ gives $\nedges\sim\mathcal{O}(500,000)$. 
This represents  a fundamental trade-off between different edge-assignment algorithms: they must simultaneously maximize \textit{efficiency}, the fraction of track segments represented as true edges, and \textit{purity}, the fraction of true edges to total edges in the hitgraph.
\mathchardef\mhyphen="2D
\begin{table*}
\centering
\caption{The $\pt$, noise, and same-layer filters are used as a handle on graph size by reducing the number of hits allowed into the graph. 
Here, we profile 100 events from the TrackML \texttt{train\_1} sample; these events have an average of $N(\mathrm{total}) = 56751\pm6070$ hits in the pixel detector. Denote the hits removed by the $\pt$, noise, and same-layer filters as $N(\pt<\pt^\mathrm{min})$, $N(\mathrm{noise})$ and $N(\mathrm{same-layer})$ respectively. 
The noise filter is observed to remove $N(\mathrm{noise})=3702\pm56$ hits, roughly 6.5\% of the detector occupancy. 
The $\pt$ and same-layer filters remove hits as a function of $\pt^\mathrm{min}$; these values are reported in the table below. 
We define $N(\mathrm{remaining})\coloneqq N(\mathrm{total})-N(\pt<\pt^\mathrm{min})-N(\mathrm{same-layer})-N(\mathrm{noise})$ to be the hits remaining after these filters are applied; $N(\mathrm{remaining})$ corresponds to $\nnodes$ constructed in the hitgraph.}
\label{tab:hit_losses}      
\begin{tabular}{ r|ll|ll }
$\pt^\mathrm{min}$ [GeV] & $N(\pt<\pt^\mathrm{min})$ & $N(\mathrm{same\mhyphen layer})$ & $N(\mathrm{remaining})$ & $N(\mathrm{remaining})/N(\mathrm{total})$ [\%] \\ \hline
2.0  & $51520 \pm 5848$ &  $439 \pm 87$  & $1090\pm156 $ & $1.9\pm0.3$ \\
1.5  & $49880 \pm 5617$ &  $921 \pm 159$ & $2248\pm155 $ & $4.0\pm0.5$ \\
1.0  & $45501 \pm 5057$ &  $2233\pm 333$ & $5315\pm152 $ & $9.4\pm1.0$ \\ 
0.9  & $43839 \pm 4855$ &  $2735\pm 395$ & $6475\pm151 $ & $11.4\pm1.2$ \\ 
0.8  & $41677 \pm 4583$ &  $3396\pm 480$ & $7976\pm150 $ & $14.1\pm1.5$ \\ 
0.7  & $38778 \pm 4242$ &  $4278\pm 582$ & $9993\pm148 $ & $17.6\pm1.9$ \\ 
0.6  & $34951 \pm 3798$ &  $5448\pm 714$ & $12650\pm146 $ & $22.3\pm2.4$ \\ 
0.5  & $29830 \pm 3265$ &  $7025\pm 875$ & $16194\pm144 $ & $28.5\pm3.1$ \\
\end{tabular}
\end{table*}

\begin{figure}[!htbp]
\centering
\includegraphics[width=0.7\columnwidth]{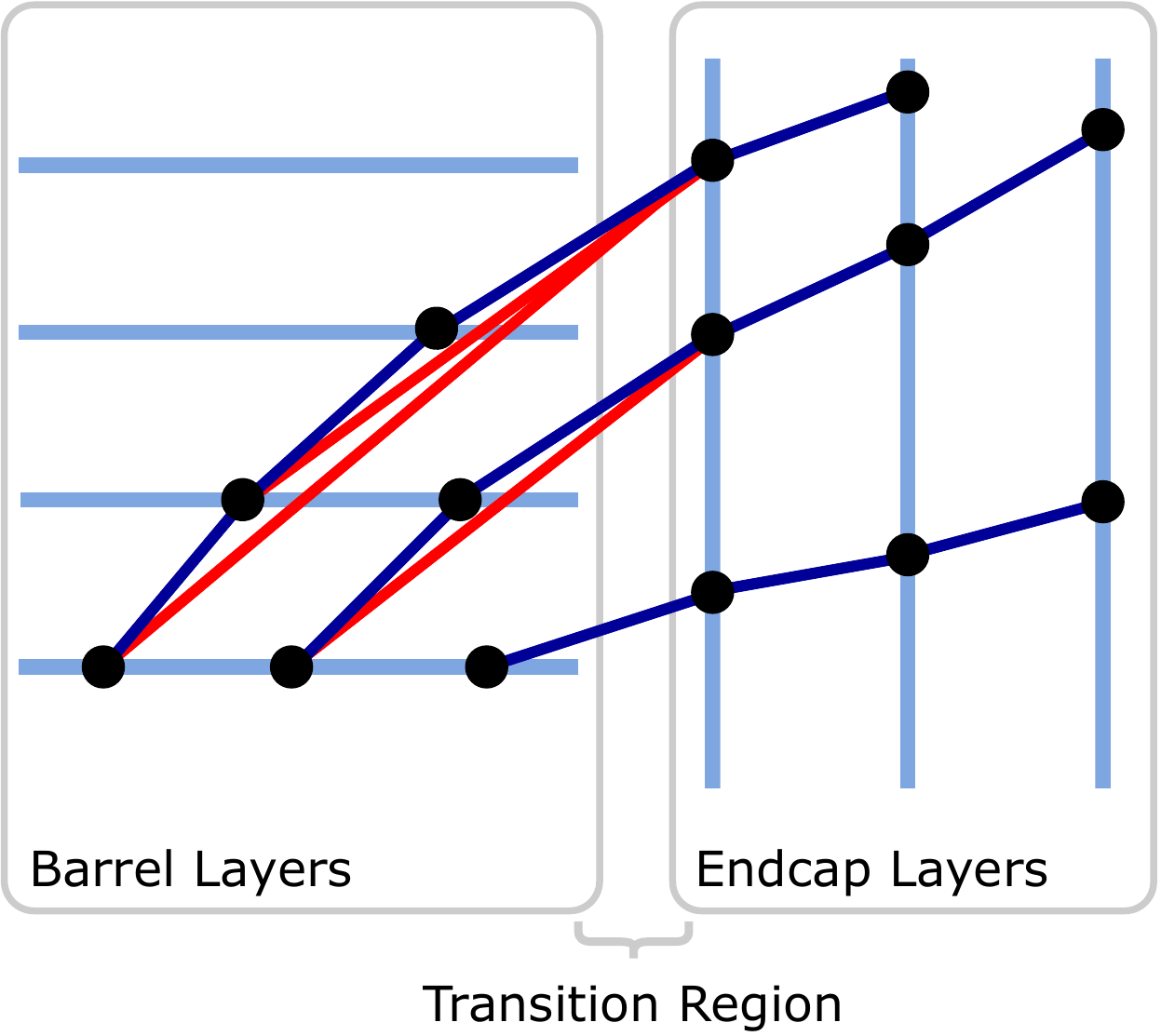}
\caption{The transition region between the barrel and endcaps introduces an ambiguity in truth-labeling edges crossing from barrel to endcap layers. 
Specifically, one may draw multiple possible edges between hits in barrel layers and the innermost endcap layer. 
Only one such edge can be true; the others (labeled red) should be rejected. 
The barrel intersection cut rejects any edges between a barrel layer and an innermost endcap layer that intersect an intermediate barrel layer. 
Accordingly, the red edges would be rejected by the intersecting line cut and the blue edges would not.}
\label{fig:intersecting_line_cut} 
\end{figure}

In this work, we compare multiple graph construction algorithms, each of which determines whether or not to extend an edge with features $a_{ij}$ between hits $i$ and $j$. 
In all methods, only pixel detector hits are considered, pseudorapidity is restricted to $\eta\in[-4,4]$, and 
the noise and same-layer hit filters are applied. 
Each method has the same definition of graph construction efficiency $(N_{\mathrm{true}}^{\mathrm{reconstructed}}/N_{\mathrm{true}}^{\mathrm{possible}})$ and purity $(N_{\mathrm{true}}^{\mathrm{reconstructed}}/N_{\mathrm{total}}^{\mathrm{reconstructed}})$.
The denominator quantity $N_{\mathrm{true}}^{\mathrm{possible}}$ is independent of the graph construction algorithm such that one may directly compare the efficiencies of the various methods.
On the other hand, the denominator $N_{\mathrm{total}}^{\mathrm{reconstructed}}$ depends on the specific graph construction routine; for this reason, it is important to study purity in the context of efficiency.
The same-layer filter introduces an ambiguity in defining edges between the barrel and innermost endcap layers. 
Specifically, barrel hits generated by the same particle could produce multiple true edges incoming to a single endcap hit. 
The resulting triangular edge pattern conflicts with the main assumption of the same-layer filter, that only one true track segment exists between each subsequent layer. 
For this reason, a \textit{barrel intersection} cut was developed, in which edges between a barrel layer and an innermost endcap layer are rejected if they intersect with any intermediate barrel layers (see Fig.~\ref{fig:intersecting_line_cut}). 

In addition to the barrel intersection cut, edges must also satisfy $\pt^\mathrm{min}$-dependent constraints on the geometric quantities $z_0=z_i-r_i\frac{z_j-z_i}{r_j-r_i}$ and $\phi_\mathrm{slope}=\frac{\phi_j-\phi_i}{r_j-r_i}$. 
These selections form the base of each of the following graph construction algorithms: 

\begin{enumerate} 
\item Geometric: Edges must satisfy the barrel intersection cut and $z_0$ and $\phi_\mathrm{slope}$ constraints. 
\item Geometric \& preclustering: In addition to all geometric selections, edges must also belong to the same cluster in $\eta$-$\phi$ space determined by the density-based spatial clustering of applications with noise (DBSCAN) algorithm~\cite{dbscan}. 
\item Geometric \& data-driven: In addition to all geometric selections, edges must connect detector modules that have produced valid track segments in an independent data sample; this data-driven strategy is known as the \textit{module map} method originally developed in ~\cite{Biscarat:2021dlj}. 
\end{enumerate}

\begin{figure}[!htbp]
\centering
\includegraphics[width=0.9\columnwidth]{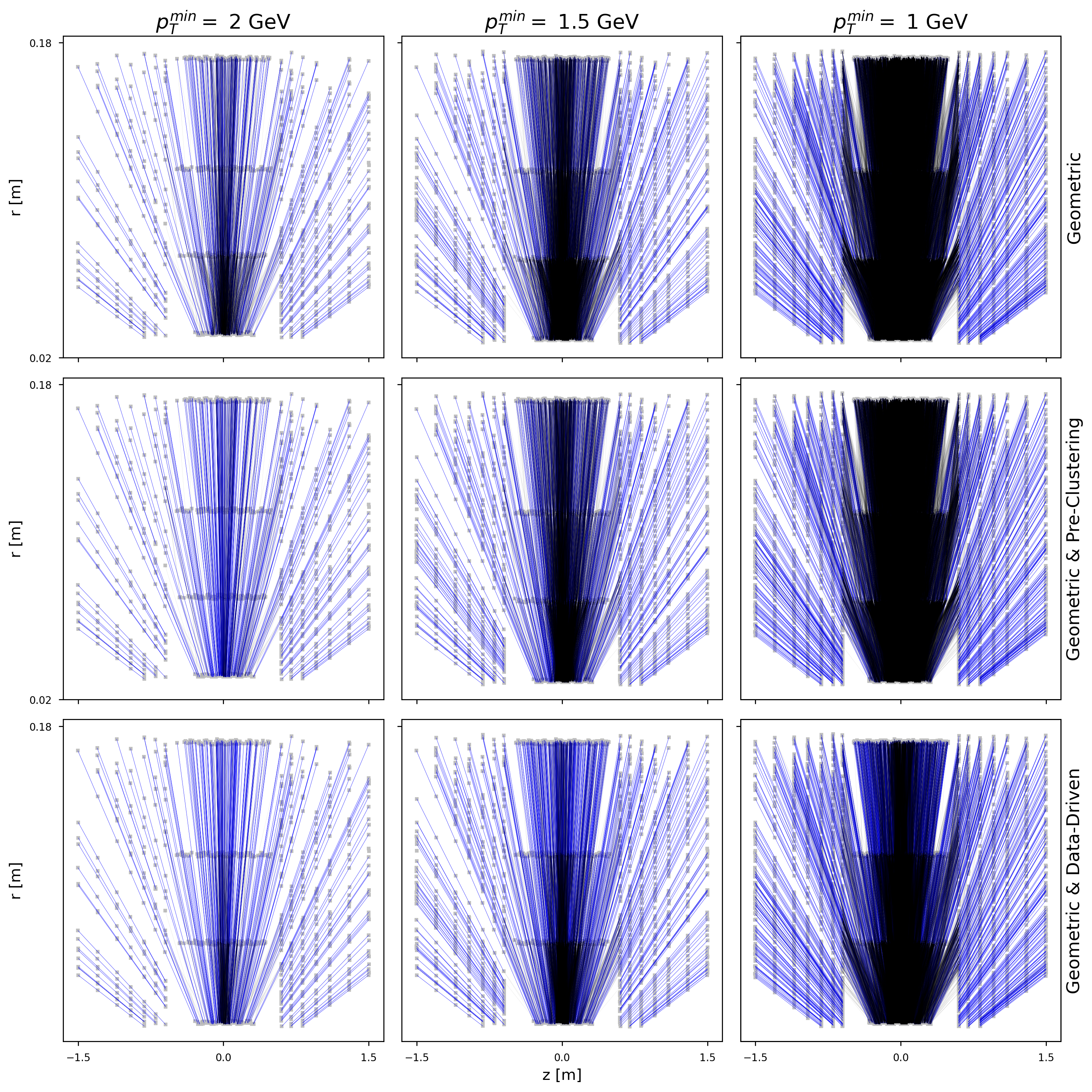}
\caption{Edge colors indicate truth labels; blue edges are true track segments and back edges are false. 
Varying $\pt^\mathrm{min}$ modulates the graph size. 
As $\pt^\mathrm{min}$ is decreased, graphs are increasingly composed of false edges. 
Preclustering and data-driven edge selections reduce the fraction of false edges in the graphs when compared to simple geometric selections. }
\label{fig:construction} 
\end{figure}

\begin{figure}[!htbp]
\centering
\includegraphics[width=\columnwidth]{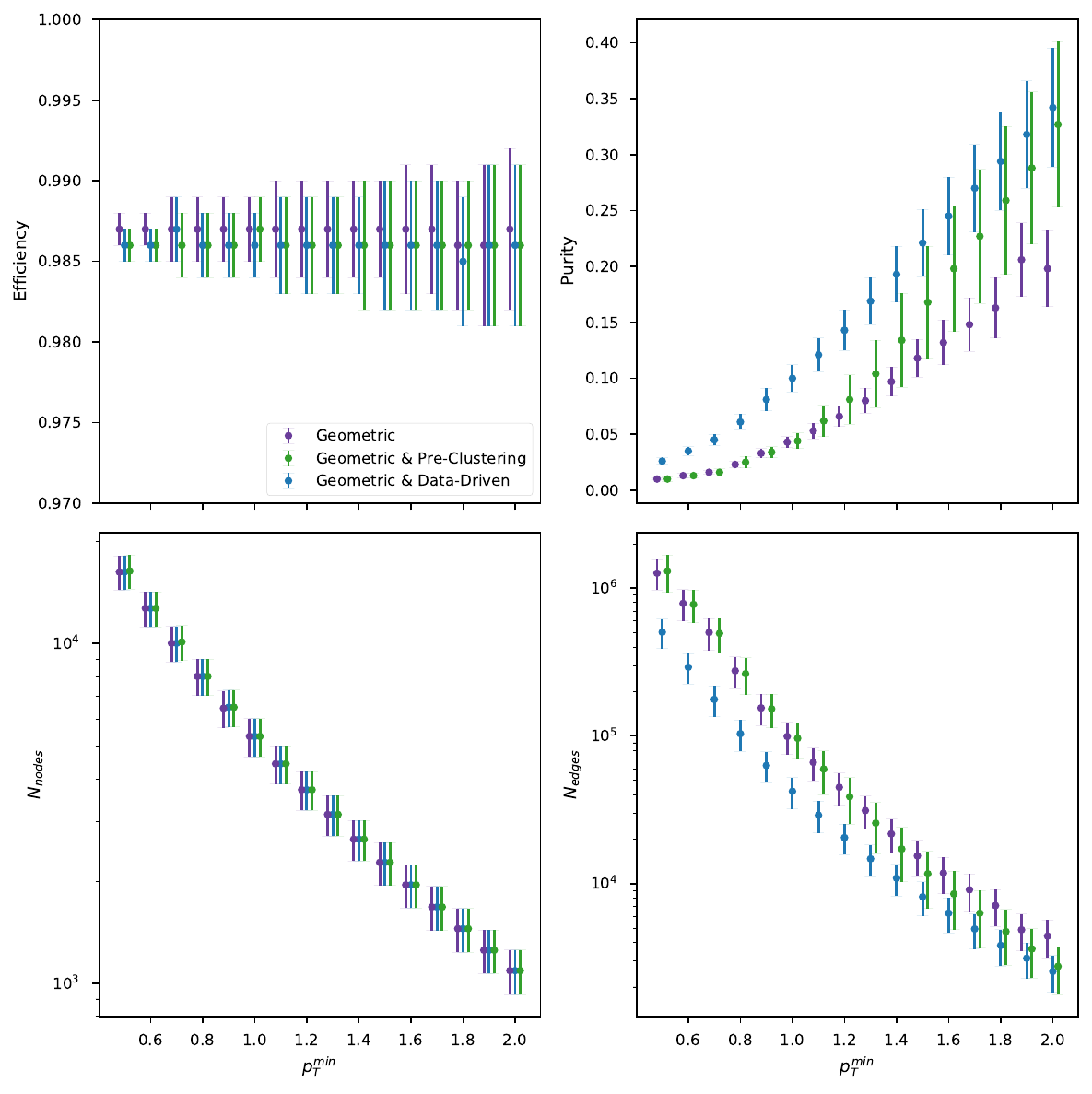}
\caption{Graph construction efficiency, purity, node counts, and edge counts are reported for a range of $\pt^\mathrm{min}$ calculated using 100 random graphs from the \texttt{train\_1} sample.}
\label{fig:build-measurements} 
\end{figure}

Truth-labeled example graphs and key performance metrics for each graph construction algorithm are shown in Figs.~\ref{fig:construction} and ~\ref{fig:build-measurements} respectively. 
For each method, $\pt^\mathrm{min}$-dependent values of $\phi_\mathrm{slope}$ and $z_0$ are chosen to keep the efficiency at a constant $\mathcal{O}(99\%)$. 
We observe a corresponding drop in purity to $\mathcal{O}(1\%)$ as $\pt^\mathrm{min}$ is decreased and graphs become denser. 
At high values of $\pt^\mathrm{min}$, preclustering hits in $\eta$--$\phi$ space yields a significant increase in purity over the purely geometric construction. 
This effect disappears as $\pt^\mathrm{min}$ decreases below $1.5\GeV$, as tracks begin to overlap non-trivially with higher detector occupancy. 
On the other hand, the data-driven module map yields a significant boost in purity for the full range of $\pt^\mathrm{min}$. 
Accordingly, the module map method is most suited to constrained computing environments in which graph size or processing time is limited. 
It should be noted, however, that purer graphs do not necessarily lead to higher edge classification accuracies. 

\subsection{Edge Classification}
\label{sec:edgeclass}

\begin{figure*}[!htbp]
\centering
  \includegraphics[width=.5\linewidth]{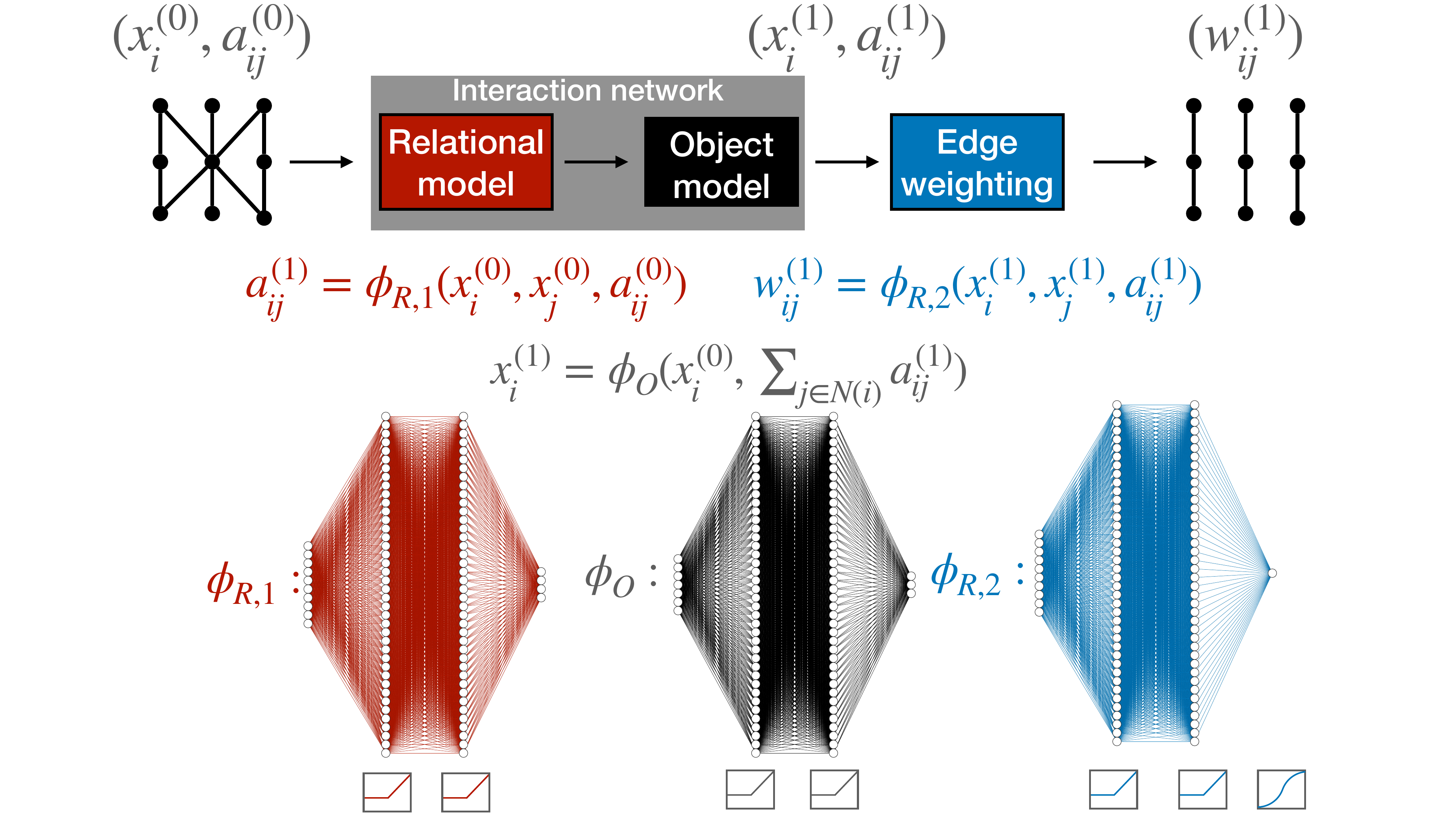}
  \includegraphics[width=.4\linewidth]{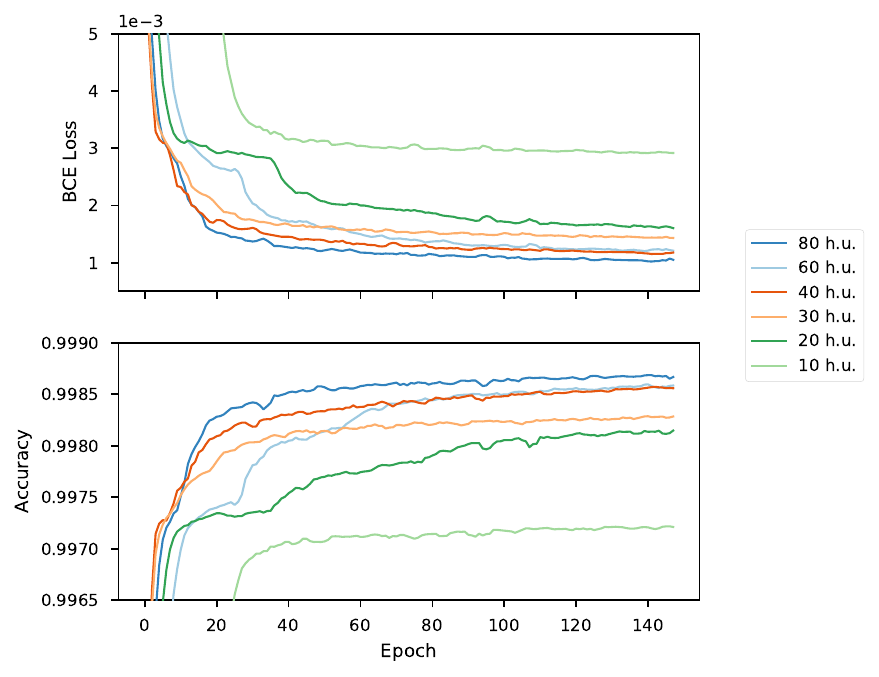}
\caption{(Left) The complete IN forward-pass with the relational and object models approximated as MLPs.
(Right) An example hyperparameter scan in which a models with varying numbers of hidden units (h.u.) were trained on $\pt^\mathrm{min}=0.7\GeV$ graphs.}
\label{fig:forward-pass}
\end{figure*}

As detailed in Sec.~\ref{sec:IN}, we have implemented the IN in \textsc{PyTorch}~\cite{pytorch} as a set of explicit matrix operations and in \textsc{PyG}~\cite{fey2019fast} as a MPNN. 
Both implementations are available in the \textsc{Git} repository accompanying this paper~\cite{IN_repo}. 
In the following studies, we limit our focus to the MPNN implementation trained on graphs built using geometric cuts only. 
Because \textsc{PyG} accommodates the sparse $\mathcal{G}_\mathrm{COO}$ edge representation, the MPNN implementation is significantly faster and more flexible than the matrix implementation (see \ref{sec:timing}). 
The full forward-pass, comprised of edge and node blocks used to predict edge weights, is shown in Fig.~\ref{fig:forward-pass}. 
The functions $\phi_{R,1}$, $\phi_{R,2}$, and $\phi_O$ are approximated as multilayer perceptrons (MLPs) with rectified linear unit (ReLU) activation functions~\cite{relu1,relu2}. 
The ReLU activation function behaves as an identity function for positive inputs and saturates at 0 for negative inputs. 
Notably, the $\phi_{R,2}$ outputs have a sigmoid activation $\sigma(\cdot)\in(0,1)$, such that they represent probabilities, or edge weights,  $W(\mathcal{G}_\mathrm{COO})\in(0,1)^{\nedges}$ that each edge is a track segment. 
We therefore seek to optimize a binary cross-entropy (BCE) loss between the truth targets $y_k=\{0,1\}$ and edge weights $w_k\in(0,1)$, which henceforth are re-labeled by the edge index $k$:
\begin{align}
    \ell\left(y_n, W_n(\mathcal{G})\right) = -\sum_{k=1}^{\nedges}\left( y_k\log w_k + (1-y_k)\log (1-w_k)\right)
\end{align}
Here, $n$ is the sample index so that the total loss per epoch is the average BCE loss $L(\{\mathcal{G}_n, y_n\}_{n=1}^N)=\frac{1}{N}\sum_{n=1}^N \ell\big(y_n, W_n(\mathcal{G})\big)$. 
Throughout the following studies, the architecture in Fig.~\ref{fig:forward-pass} is held at a constant size of 6,448 trainable parameters, corresponding to 40 hidden units (h.u.) per layer in each of the MLPs.
Validation studies indicate that even this small network rapidly converged to losses of $\mathcal{O}(10^{-3})$, similar to its larger counterparts (see Fig.~\ref{fig:forward-pass}). 
Assuming every MLP layer has the same number of h.u., 40 h.u. per layer is sufficient to recover the maximum classification accuracy with models trained on $\pt^\mathrm{min}=1\GeV$ graphs.
In the following studies, models are trained on graphs built with $\pt^\mathrm{min}$ ranging from 0.6--2\GeV. 
At each value of $\pt^\mathrm{min}$, 1500 graphs belonging to the TrackML \texttt{train\_1} sample are randomly divided into 1000 training, 400 testing, and 100 validation sets. 
The Adam optimizer is used to facilitate training~\cite{adam}. 
It is configured with learning rates of 3.5--8$\times10^{-3}$, which are decayed by a factor of $\gamma=0.95$ for $\pt^\mathrm{min}\leq 1\GeV$ and $\gamma=0.8$ for $\pt^\mathrm{min}> 1\GeV$ every 10 epochs.

In order to evaluate the IN edge-classification performance, it is necessary to define a threshold $\delta$ such that each edge weight $w_k\in W(\mathcal{G}_{COO})$ satisfying $w_k\geq\delta$ or $w_k<\delta$ indicates that edge $k$ was classified as true or false respectively. 
Here, we define $\delta^*$ as the threshold at which the true positive rate (TPR) equals the true negative rate (TNR). 
In principle, $\delta^*$ may be calculated individually for each graph. 
However, this introduces additional overhead to the inference step, which is undesirable in constrained computing environments. 
We instead determine $\delta^*$ during the training process by minimizing the difference $|\mathrm{TPR}-\mathrm{TNR}|$ for graphs in the validation set. 
The resulting $\delta^*$, which is stored for use in evaluating the testing sample, represents the average optimal threshold for the validation graphs. 
Accordingly, we define the model's accuracy at $\delta^*$ as $(n_\mathrm{TP}+n_\mathrm{TN})/\nedges$, where $n_\mathrm{TP}$  ($n_\mathrm{TN}$) is the number of true positives (negatives), and note that the BCE loss is independent of $\delta^*$. 

\begin{figure*}[!htbp]
\centering
  \includegraphics[width=.45\linewidth]{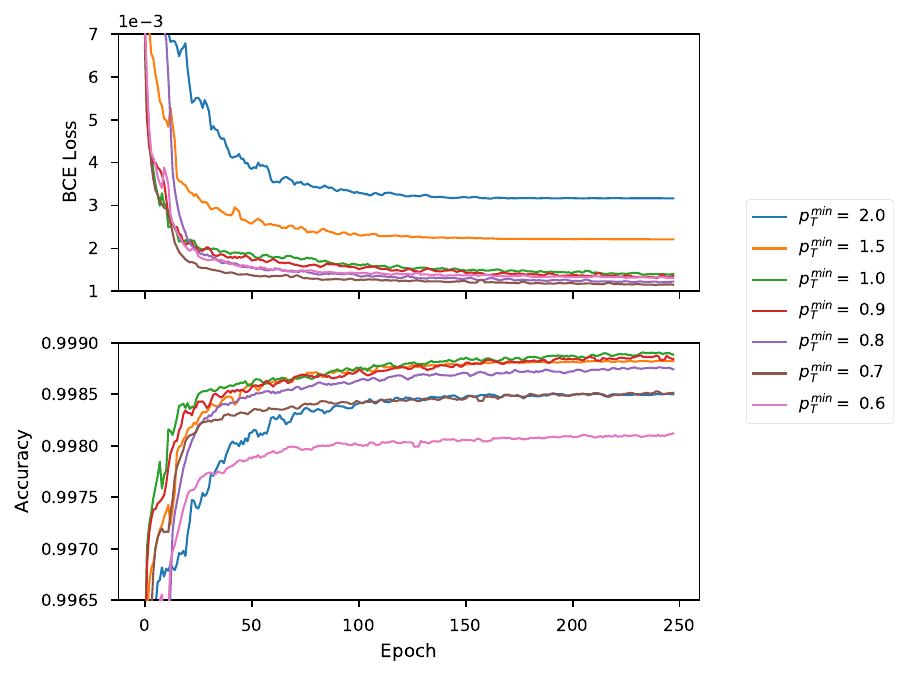}
  \includegraphics[width=.5\linewidth]{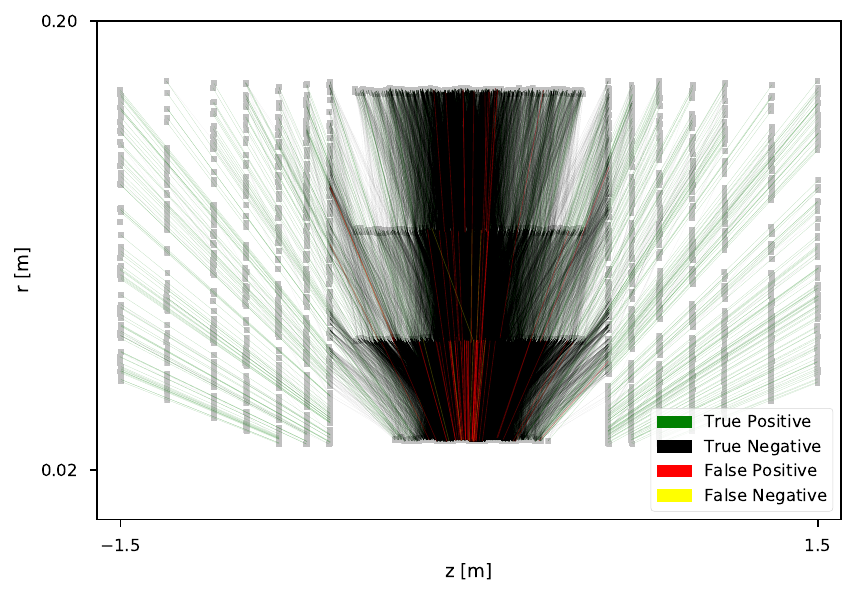}
\caption{(Left) Loss convergence for models trained on various $\pt^\mathrm{min}$ graphs. 
(Right) A model trained on $\pt^\mathrm{min}=1\GeV$ graphs was used to evaluate an unseen $\pt^\mathrm{min}=1\GeV$ graph, yielding a loss of $1.52\times10^{-3}$ and accuracy of 99.9\%. 
98 out of 95,160 edges were incorrectly classified; these erroneous classifications are magnified in the figure.}
\label{fig:classification}
\end{figure*}

\begin{figure}[!htbp]
\centering
\includegraphics[width=\columnwidth,clip]{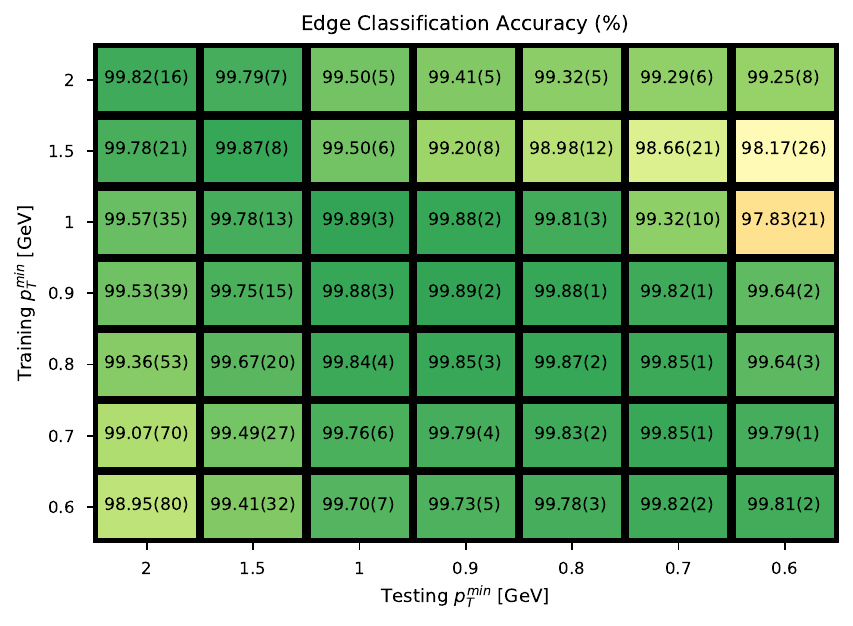}
\caption{Models trained on various $\pt^\mathrm{min}$ graphs in the \texttt{train\_1} sample were tested on 400 graphs from the \texttt{train\_3} sample at various $\pt^\mathrm{min}$ thresholds. }
\label{fig:accuracy}
\end{figure}

As shown in Fig.~\ref{fig:classification}, the training process results in smooth convergence to excellent edge-classification accuracy for a range of $\pt^\mathrm{min}$. 
Classification accuracy degrades slightly as $\pt^\mathrm{min}$ is lowered below $1\GeV$; hyperparameter studies indicate that larger networks improve performance on lower $\pt^\mathrm{min}$ graphs (see Fig.~\ref{fig:forward-pass}). 
A transfer learning study was conducted in which models trained on graphs at a specific $\pt^\mathrm{min}$ were tested on graph samples at a range of $\pt^\mathrm{min}$. 
The results are summarized in Fig.~\ref{fig:accuracy}, which shows that the models achieve relatively robust performance on a range of graph sizes. 
These results suggest it may be possible to train IN models in simplified scenarios and apply them to more complex realistic scenarios (e.g. without a $\pt^\mathrm{min}$ cut).

\subsection{Track Building}
\label{sec:trackbuilding}

In the track building step, the predicted edge weights $w_k\in W(\mathcal{G}_\mathrm{COO})$ are used to infer that edges satisfying $w_k\geq\delta^*$ represent true track segments. 
If the edge weight mask perfectly reproduced the training target (i.e. $\mathtt{int}(W(\mathcal{G}_\mathrm{COO})\geq\delta^*)=y$), the edge-classification step would produce $n_\mathrm{particles}$ disjoint subgraphs, each corresponding to a single particle. 
Imperfect edge-classification leads to spurious connections between these subgraphs, prompting the need for more sophisticated track-building algorithms. 
Here, we use the union-find algorithm~\cite{unionfind} and DBSCAN to cluster hits in the edge-weighted graphs. 
Hit clusters are then considered to be reconstructed tracks candidates; the track candidates are subsequently matched to simulated particles (when possible). 
In a full tracking pipeline, these track candidates would then be fit to extract track parameters such as $\pt$ and $\eta$; in this work we use truth information for matched particles to get the track parameters.
Tracking efficiency metrics measure the relative success of the clustering and matching process using various definitions. 
We define three tracking efficiency measurements using progressively tighter requirements to allow comparison with current tracking algorithm efficiencies and other on-going HL-LHC tracking studies:
\begin{enumerate}
    \item LHC match efficiency: the number of reconstructed tracks containing over 75\% of hits from the same particle, divided by the total number of particles.
    \item Double-majority efficiency: the number of reconstructed tracks containing over 50\% of hits from the same particle and over 50\% of that particle's hits, divided by the total number of particles.
    \item Perfect match efficiency: the number of reconstructed tracks containing only hits from the same particle and every hit generated by that particle, divided by the number of particles.
\end{enumerate}
We note that the perfect match efficiency is not commonly used by experiments as 100\% is not realistically achievable, but we present it to demonstrate the absolute performance of the GNN tracking pipeline.

\begin{figure*}[!htbp]
\centering
\includegraphics[width=2\columnwidth,clip]{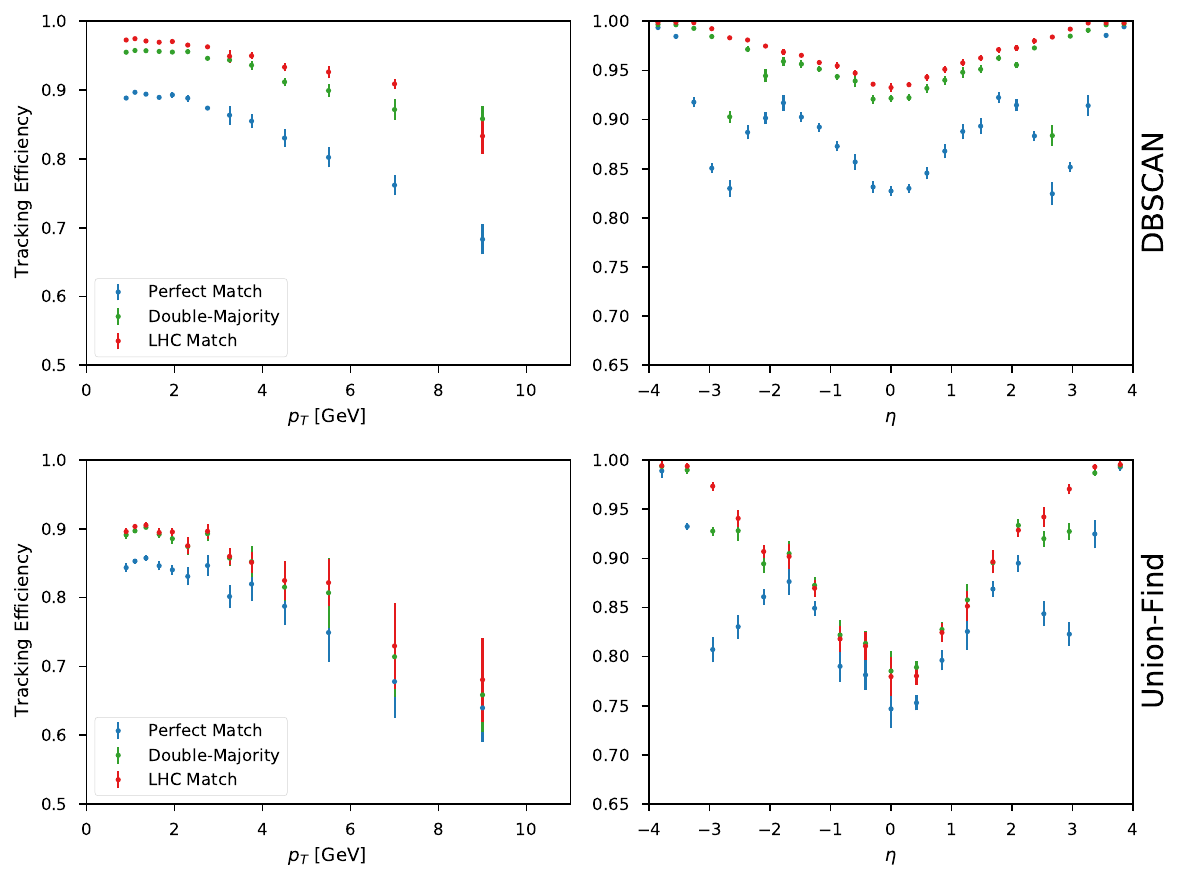}
\caption{The track-building performance of DBSCAN and union-find are measured as a function of particle $\pt$ and $\eta$ for three tracking efficiency definitions at $\pt^\mathrm{min}=0.9\GeV$. }
\label{fig:tracking-effs}
\end{figure*}

Figure~\ref{fig:tracking-effs} shows each of these tracking efficiencies as a function of particle $\pt$ and $\eta$ for both the DBSCAN and union-find clustering approaches. 
Additionally, Table~\ref{tab:fake-rate} shows the corresponding fake rates, or fractions of unmatched clusters relative to all clusters, across the full $\pt$ and $\eta$ range. 
The efficiencies and fake rates are calculated with $\pt^\mathrm{min}=0.9\GeV$ graphs.
Tracking performance is relatively stable at low $\pt$ but degrades for higher $\pt$ particles; similar effects have been noted in other edge-weight-based hit clustering schemes \cite{Biscarat:2021dlj}. 
The tracking efficiencies are lowest in the neighborhood of $\eta=0$, indicating that performance is worst in the pixel barrel region.
This is consistent with the observation that most edge classification errors occur in the barrel, where the density of detector modules is significantly higher \cite{TrackML}. 
Tracking efficiency loss around $|\eta|\approx 2.5$ corresponds to the transition region between barrel and endcap layers. 
DBSCAN demonstrates higher tracking efficiency than union-find across all $\pt$ and $\eta$ values and efficiency definitions. 
This performance gap is likely due to the additional spatial information used in DBSCAN's clustering routine. 
Moving forward, additional tracking performance may be recovered by leveraging the specific values of each edge weight to make dynamic hit clustering decisions. The fake rates are relatively low for both track-building methods, and as expected roughly increase for increasingly tight efficiency definitions. 
Interestingly, DBSCAN demonstrates a lower fake rate for LHC match efficiency while union-find demonstrates a lower fake rate for the perfect match efficiency; DBSCAN also has a larger drop in tracking efficency between the double match and perfect match definitions, indicating that while DBSCAN identifies more track candidates, union-find builds tracks more precisely.

\begin{table}[!htbp]
    \centering
    \begin{tabular}{r|ll}
        Efficiency definition & Union-find & DBSCAN\\
        \hline
        LHC match & $0.0471\pm0.008$ & $0.0275\pm0.005$\\
        Double majority & $0.0934\pm0.01$ & $0.0891\pm0.01$\\
        Perfect match & $0.0910\pm0.01$ & $0.1242\pm0.01$\\
    \end{tabular}
    \caption{Overall fake rates of union-find and DBSCAN track-building for three tracking efficiency definitions for $\pt^\mathrm{min}=0.9\GeV$.}
    \label{tab:fake-rate}
\end{table}

\subsection{Inference Timing}
\label{sec:timing}

An important advantage of GNN-based approaches over traditional methods for HEP reconstruction is the ability to natively run on highly parallel computing architectures.
The \textsc{PyG} library supports graphics processing units (GPUs) to parallelize the algorithm execution.
Moreover, the model was prepared for inference by converting it to a TorchScript program~\cite{torchscript}.
For the IN studied in this work, the average CPU and GPU inference times per graph for a variety of minimum $\pt$ cuts is shown in Table~\ref{tab:timing}.
For this test, the graphs are constructed using the geometric selections as described in Section~\ref{sec:Graph}. 
Moreover, we use bidirectional graphs, which means both directed edges (outward and inward from the primary vertex) are present in the edge list.
As can be seen, inference can be significantly sped up with heterogeneous resources like GPUs.
For instance, for a 0.5\GeV minimum \pt cut, the inference time can be reduced by approximately a factor of 10 using the GPU with respect to the CPU.
In general, the speedup is greater at lower $\pt^\mathrm{min}$ because of the higher multiplicity and thus the greater gain from parallelization on the GPU versus the CPU.
Other heterogeneous computing resources specialized for inference may be even more beneficial. 
This speed up may benefit the experiments' computing workflows by accessing these resources as an on-demand, scalable service~\cite{Krupa:2020bwg,Rankin:2020usv,Wang:2020fjr}.

\begin{table*}
\centering
\caption{CPU and GPU inference time estimates for each $\pt$ threshold. 
The model was prepared for inference by converting it to a TorchScript program.
The timing test were performed with an Nvidia Titan Xp GPU with 12\,GB RAM and a 12-core Intel Xeon CPU E5-2650 v4 @ 2.20\,GHz.
Inference is performed with a batch size of one graph. 
Graphs are constructed using geometric restrictions with bidirectional edges (both edge directions are present).
The inference is repeated 100 times (after some warm-up) for 5 iterations and the best time per inference over the 5 iterations is found.
The mean and standard deviation of the best inference time derived for 5 random graphs in the testing dataset is then reported.
The mean and standard deviation of the number of nodes and edges are also reported for 100 graphs.
We find a significant speedup with the GPU versus the CPU, which is greater at lower $\pt^\mathrm{min}$ because of the higher multiplicity and thus the greater gain from parallelization on the GPU.}
\label{tab:timing}      
\begin{tabular}{ r|llll }
$\pt^\mathrm{min}$ [GeV] & CPU [ms] & GPU [ms] & $\overline{n}_\mathrm{nodes}$ & $\overline{n}_\mathrm{edges}$ \\\hline
 2  &  $3.83 \pm 0.89$ &  $0.95 \pm 0.01$ & $1090.6 \pm 192.8$ & $9080.7 \pm 3027.1$ \\
 1.5  & $7.96 \pm 1.44$ & $0.95 \pm 0.03$ & $2247.0 \pm 363.9$ & $30980.6 \pm 9468.6$\\
  1  & $33.96 \pm 11.24$ & $3.61 \pm 0.91$ & $5309.3 \pm 765.5$ & $200910.1 \pm 55825.9$ \\
 0.9 & $52.44 \pm 14.15$ & $5.36 \pm 1.37$ & $6468.5 \pm 912.2$ & $312809.5 \pm 85441.3$ \\
 0.8 & $91.86 \pm 24.42$ & $9.60 \pm 2.61$ & $7970.5 \pm 1100.0$  &  $556417.7 \pm 151482.8$\\
 0.7 & $168.40 \pm 41.34$ & $17.70 \pm 4.39$ &  $9982.7 \pm 1341.5$ & $1011884.4 \pm 268706.0$\\
 0.6  & $273.20 \pm 62.09$ & $28.84 \pm 6.65$ & $12640.0 \pm 1648.2$ & $1585883.3 \pm 409146.8$ \\
 0.5  & $437.00 \pm 97.99$ &  $44.66 \pm 7.91$& $16178.6 \pm 2019.1$ &  $2535979.6 \pm 628297.1$\\
\end{tabular}
\end{table*}

Work has also been done to accelerate the inference of deep neural networks with heterogeneous resources beyond GPUs, like field-programmable gate arrays (FPGAs)~\cite{FINN,FINNR,fpgadeep,fpgaover,Duarte:2018ite,Summers:2020xiy,bnnpaper,Coelho:2020zfu,Aarrestad:2021zos}.
This work extends to GNN architectures~\cite{Iiyama:2020wap,IN_fpga}.
Specifically, in Ref.~\cite{IN_fpga}, a compact version of the IN was implemented for $\pt > 2\GeV$ segmented geometric graphs with up to 28 nodes and 37 edges, and shown to have a latency less than 1\,$\mu$s, an initiation interval of 5\,ns, reproduce the floating-point precision model with a fixed-point precision of 16 bits or less, and fit on a Xilinx Kintex UltraScale FPGA. 

While this preliminary FPGA acceleration work is promising, there are several limitations of the current FPGA implementation of the IN:
\begin{enumerate}
    \item This fully-pipelined design cannot easily scale to beyond $\mathcal O(100)$ nodes and $\mathcal O(1,000)$ edges. However, if the initiation interval requirements are loosened, it can scale up to $O(10,000)$ nodes and edges.
    \item The neural network itself is small, and while it is effective for $\pt> 2\GeV$ graphs, it may not be sufficient for lower-$\pt$ graphs.
    \item The FPGA design makes no assumptions about the possible graph connectivity (e.g. layer 1 nodes are only connected to layer 2 nodes), and instead allows all nodes to potentially participate in message passing. 
    However by taking this additional structure into account, the hardware resources can be significantly reduced.
    \item Quantization-aware training~\cite{bertmoons,NIPS2015_5647,zhang2018lq,ternary-16,zhou2016dorefa,JMLR:v18:16-456,xnornet,micikevicius2017mixed,Zhuang_2018_CVPR,wang2018training,bnnpaper} using \textsc{QKeras}~\cite{Coelho:2020zfu,qkeras} or \textsc{Brevitas}~\cite{FINNR,brevitas}, parameter pruning~\cite{optimalbraindamage,han2016deep,lotteryticket,supermask,stateofpruning,Hawks:2021ruw}, and general hardware-algorithm codesign can significantly reduce the necessary FPGA resources by reducing the required bit precision and removing irrelevant operations.
    \item The design can be made more flexible, configurable, and reusable by integrating it fully with a user-friendly interface like \texttt{hls4ml}~\cite{hls4ml}.
\end{enumerate}

\section{Summary and Outlook}
\label{sec:summary}
In this work, we have shown that the physics-motivated interaction network (IN), a type of graph neural network (GNN), can successfully be applied to the task of charged particle tracking across a range of hitgraph sizes. 
Through a suite of graph construction, edge classification, and track building measurements, we have framed the IN's performance in the context of a GNN-based tracking pipeline following a truth-based hit filtering preselection in which hits associated with particles whose transverse momentum ($\pt$) is below a certain threshold ($\pt^\mathrm{min}$) are removed. 
The graph construction measurements demonstrate that geometric cuts, hit clustering, and data-driven strategies are effective in constructing highly-efficient graphs from pixel barrel and endcap layers; in constrained computing environments, the parameters of each strategy allow a trade-off between graph efficiency and purity. 
In particular, for a fixed graph construction efficiency of $\mathcal{O}(99\%)$, we show that geometric cuts alone produce reasonably pure graphs ($\sim4\%$ purity at $\pt^\mathrm{min}=1\GeV$) but that the module-map method produces the most pure graphs for the entire range of $\pt^\mathrm{min}$ ($\sim10\%$ purity at $\pt^\mathrm{min}=1\GeV$).
With efficiency held constant, purity is more-or-less a comparison of graph sizes, indicating that the module map method is most suited for graph construction in constrained computing environments. 
Though high graph efficiency is desirable in a global sense, graph purity is non-trivially related to downstream physics performance; in particular, many message passing GNN architectures may benefit from less-pure graphs due to higher edge connectivity.

The lightweight IN models trained in the edge classification step demonstrate extremely high edge classification efficiency for a range of $\pt^\mathrm{min}$.
Significantly, we find models trained in simpler scenarios (larger $\pt^\mathrm{min}$) generalize to more complex scenarios (smaller $\pt^\mathrm{min}$).
Track building measurements performed on these edge-weighted graphs showed that DBSCAN's spatial clustering outperformed union-find clustering across a variety of efficiency definitions.

The IN architecture presented here is substantially smaller than previous GNN tracking architectures, which may enable its use in constrained computing environments.
Accordingly, we have compared the IN's CPU and GPU inference times and discussed related work on accelerating INs with FPGAs. 
As described in Section~\ref{sec:timing}, there are several limitations to the current FPGA implementation of the IN and addressing these concerns is the subject of ongoing work. 
  

Another important aspect of GNN-based tracking is reducing the time it takes to construct graphs. 
Ongoing efforts are dedicated to studying how best to accelerate graph construction using heterogeneous resources.
Alternative GNN approaches that do not require an input graph structure, such as dynamic graph convolutional neural networks~\cite{DGCNN}, distance-weighted GNNs~\cite{Qasim:2019otl}, attention-based transformers~\cite{vaswani2017attention}, reformers~\cite{kitaev2020reformer}, and performers~\cite{choromanski2021rethinking}, may be fruitful avenues of investigation as well.

In summary, geometric deep learning methods can be naturally applied to many physics reconstruction tasks, and our work and related studies establish GNNs as an extremely promising candidate for tracking at the high luminosity LHC. 




\begin{acknowledgements}
We gratefully acknowledge the input and discussion from the Exa.TrkX collaboration.
\end{acknowledgements}


\bibliographystyle{cms_unsrt}       
\bibliography{refs}   
\end{document}